\documentclass{article}

\usepackage{arxiv}

\usepackage[utf8]{inputenc} 
\usepackage[T1]{fontenc}    
\usepackage{hyperref}       
\usepackage{url}            
\usepackage{booktabs}       
\usepackage{amsfonts}       
\usepackage{nicefrac}       
\usepackage{microtype}      

\usepackage{epsfig}
\usepackage{epstopdf}
\usepackage{amsmath}

\newcommand{\bs}{\boldsymbol}

\title{Beams with propagation-invariant transverse polarization pattern}  
\author{ J. C. G. de Sande\\
ETSIS de Telecomunicaci\'{o}n \\ Universidad Polit\'{e}cnica de Madrid \\ Campus Sur 28031 Madrid, Spain\\
    \texttt{juancarlos.gonzalez@upm.es} \\
   \And
 Gemma Piquero \\
  Departamento de \'{O}ptica\\ Universidad Complutense de Madrid\\ Ciudad Universitaria,  28040 Madrid, Spain\\
  \texttt{piquero@ucm.es} \\
  \AND
 Juan Carlos Suarez-Bermejo\\
 Materials Science\\ Universidad Polit\'{e}cnica de Madrid\\ Av. de la Memoria 4, 28040 Madrid, Spain\\
    \texttt{juancarlos.suarez@upm.es} 
   \And
   Massimo Santarsiero\\
   Dipartimento di Ingegneria\\ Universit\`a Roma Tre\\ Via V. Volterra 62, 00146 Rome, Italy\\
   \texttt{massimo.santarsiero@uniroma3.it} \\
}

\begin{document}
\maketitle

\begin{abstract}
A wide class of nonuniformly totally polarized beams is introduced that preserve their transverse polarization pattern during paraxial propagation. They are obtained as suitable combinations of Gaussian modes and find applications in polarimetric techniques that use a single input beam for the determination of the Mueller matrix of a homogeneous sample.
The class also includes beams that present all possible polarization states across their transverse section (Full-Poincar\'e beams). An example of such beams and its use in polarimetry is discussed in detail. The requirement of invariance of the polarization pattern can be limited to the propagation in the far field, in which case less restrictive conditions are found and a wider class of beams is obtained. 
\end{abstract}


Polarimetry is a noninvasive testing technique that provides information about the optical properties of a sample. Since partially polarized light can be described by the $4 \times 1$ Stokes vector $\bf{S}$, the optical behavior of a sample can be described by its $4 \times 4$ Mueller matrix, $\widehat M$~\cite{chipman18, Goldstein03, Gil16}. To measure its Mueller matrix, a sample is generally tested sequentially by at least four different input polarization states~\cite{chipman18}.  The latter must be represented by four independent $4 \times 1$ Stokes vectors, in order for the Mueller matrix elements to be recovered through an inversion procedure~\cite{chipman18, Goldstein03}. 

A polarimetric technique has been recently proposed  for the analysis of linear, deterministic and homogeneous samples~\cite{Sande:OLEN17}. It uses a single test beam, but endowed with nonuniform polarization across its transverse profile, and the response of the sample is then analyzed at different points of the transverse beam profile. To this aim, beams that potentially provide the best performances are the ones that present all possible polarization states across their profile, because using a large number of input polarization states results in a reduction of the uncertainties on the measured values of the Mueller matrix~\cite{SUAREZBERMEJO:OLEN19}. Beams of this type were introduced by Beckley {\it et al}~\cite{Beckley:OE10} and are known as Full-Poincar\'e Beams (FPBs)~\cite{Galvez:AO12,Alieva:JOSAA13,Alpmann:SR17,Li:JAP19}. We shall use the term Full-Poincar\'e Polarimetry (FPP) to denote this technique. 

In previous experimental verifications of FPP, the input FPB was obtained on exploiting the anisotropy properties of a calcite crystal~\cite{SUAREZBERMEJO:OLEN19,Piquero:JOPT18}. The transverse dimensions of the tested samples were large enough to allow the use of an input FPB with spot size of the order of some millimeters, so that no optical magnifying systems were needed to image the output beam onto a CCD sensor and perform measurements.

For smaller samples, or when their optical properties are not uniform, the input beam needs to be focused onto the region to be tested. The analysis of the output beam would necessarely require the use of optical systems, which could affect the polarization measurements. Therefore, it would be convenient to have a FPB that maintains its polarization profile during free propagation, except for a transverse scaling factor, so that the same pattern would be observed at any distance from the sample. The most comfortable beam size for the detection could be selected on simply choosing the right distance from the sample.

In this paper we first present a class of perfectly coherent and non-uniformly totally polarized (NUTP) beams~\cite{Piquero:JOSAA20} that fulfill the above requirements, that is, they retain the shape of their transverse polarization pattern during paraxial propagation, the only change being a scaling of the transverse coordinates as one goes from one transverse plane to another. To the same class belongs all beams that are obtained by transforming one of the above beams by means of a general linear, deterministic and homogeneous optical element. In other terms, if the input beam fulfills the above requirement, the propagation invariance is guaranteed for the beam exiting the sample, too. Within such class, we will identify NUTP beams that are FPB, and will study their applicability in FPP. Conditions ensuring the polarization invariance in propagation has been studied for the case of uniformly polarized, partially coherent electromagnetic fields~\cite{Wolf:OL07,Korotkova:OL11,RMH:OC07}. Here we are interested to perfectly polarized and perfectly coherent fields. 

In this case, a condition ensuring the shape-invariance of the transverse polarization pattern of a field is that the latter be expressible as the superposition of two component fields, having orthogonal and uniform polarizations, chosen in such a way that the ratio between their values at any point of a transverse plane (taking the above scaling factor into account) remains invariant at any propagation distance $z$. In the trivial case of uniform polarization this happens because one of the two fields can always be set to zero.
 So, we denote by ${\bs V}_{0}({\bs r}) $ a typical vector field across the plane $z=0$, write it as
\begin{equation}
{\bs V}_{0}({\bs r}) 
= 
\displaystyle\sum_{i=1}^2
V^{(i)}_{0} ({\bs r})  \, {\hat \epsilon}_i
\; ,
\label{super}
\end{equation}
where ${\hat \epsilon}_i$ $(i=1,2)$ are unit vectors specifying two orthogonal polarization states,  and look for  two component fields, $V^{(i)}_{0} ({\bs r})$, fulfilling the above condition. 

Within the paraxial approximation, the field propagated at any distance $z$ can be evaluated by expressing the initial one in terms of Gaussian modes, whose propagation expression can be given in closed form. This can be made both in rectangular or in polar coordinates, where Hermite--Gaussian (HG) or Laguerre--Gaussian (LG) functions, respectively, can be used~\cite{ABRAMOCHKIN:OC91}. We denote both families of functions by $\Psi_{h}({\bs r}; w_0)$, regardless of the specific coordinate system we are using, while the meaning of the index $h$  depends on the particular class of functions: it may represent either the indexes (say, $n$ and $m$) of the two Hermite polynomials involved in the definition of HG functions, or the two parameters (say, $p$ and $s$) specifying a generalized Laguerre polynomial, for the case of LG modes. Both classes of functions depend on a further parameter, namely, the spot size $w_0$, which fixes the transverse extent of the modes at their waist.

Gaussian functions form a complete set in $L^2$. This allows for any paraxial beam to be written as a superposition of modes of the above types, with arbitrary $w_0$. In particular, we have
\begin{equation}
\label{sviluppoHGin0}
V^{(i)}_0({\bs r})
=
\displaystyle\sum_{h}
c^{(i)}_{h}
\,
{\Psi}_{h}({\bs r}; w_0)
\; ,
\end{equation}
where
\begin{equation}
c^{(i)}_{h}
=
\displaystyle\int\limits
V^{(i)}_0({\bs r})
\;
{\Psi}^*_{h}({\bs r}; w_0)
\; {\rm d}{\bs r}
\; ,
\label{coefficientiHG}
\end{equation}
the integral being extended to the whole plane $z=0$. The field propagated at the distance $z$ is therefore evaluated taking into account the effects of the propagation on each of the modes. The latter, in fact, keep their initial form, up to a scaling factor, and acquire both a spherical curvature and a phase depending on the mode indices. More precisely,  we have
\begin{equation}
\begin{array}{rl}
V^{(i)}_z({\bs r})
=
&
\displaystyle\frac{w_0}{w_z}
\; e^{{\rm i}kz}
\exp\left(\frac{{\rm i}k r^2}{2 R_z}\right)
\displaystyle\sum_{h}
\; c^{(i)}_{h}
\,
{\Psi}_{h}({\bs r}; w_z)
\; e^{- \displaystyle {\rm i} (N+1) \Phi_z}
\; ,
\end{array}
\label{sviluppoHGinZ}
\end{equation}
where $k$ is the wave number and
\begin{eqnarray}
\begin{array}{c}
w_z
=  \hspace{-2mm} 
w_0
\sqrt{1 + \left(\frac{z}{z_R}\right)^2}
; \,\,\, \,\,\,
R_z
=
z
\left[1 + \left(\frac{z_R}{z}\right)^2\right]; \,\,\, \,\,\,
 \Phi_z
=
\arctan
\left(\frac{z}{z_R}\right) 
, 
\end{array}
\label{parameters}
\end{eqnarray}
with $z_R=k w_0^2/2$ being the Rayleigh distance. The parameter $N$ is related to the indexes of the modes. It equals $n+m$ for HG modes and $2 p + |s|$ for LG modes and contributes to the phase term $(N+1) \Phi_z$ appearing in Eq.~(\ref{sviluppoHGinZ}). Such term is known as \emph{phase anomaly} or \emph{Gouy phase}~\cite{Siegman:Lasers86}. It is important to stress that, since each of the modes is shape-invariant during propagation, the phase anomaly is the only responsible for the fact that a general beam changes in shape during propagation.

According to Eq.~(\ref{super}), the polarization pattern of a beam across a transverse plane is determined by the relative amplitude and phase of the two component fields at any point of the plane and, in general, changes during propagation because the two fields generally change in shape and acquire different phases. 
On the other hand, from the above results, it turns out that a condition ensuring that the relative amplitudes and phases of the two beams remain unchanged at any point is to require that, in the modal expansion of its component fields [Eqs.~(\ref{sviluppoHGinZ})], only those terms are present for which the phase anomaly changes in the same way during propagation. This means that either of the two orthogonally polarized components of the beam has to be expressed as the sum of HG (and/or LG) modes with one and the same value of $N$. This condition implies that, except for overall amplitude and phase terms, the propagated field is an exact replica of the initial one, scaled by the factor $w_0/w_z$. 

It is worth recalling here that the LG mode with indices $p$ and $s$ can be expressed as the sum of HG modes for which $n+m=2p+|s|$~\cite{Allen:PRA92,ABRAMOCHKIN:OC91}  
and, conversely, the HG mode with indices $n$ and $m$ can be expressed as the sum of LG modes for which $2p+|s|=n+m$. This means that every beam obtained following the above rule can equally be thought of as a superposition of either HG or LG modes, the involved expansion coefficients being related by linear transformation rules~\cite{O'Neil:OC00}. 

A significant remark can be made before dealing with a particular example. If a beam presents a propagation-invariant polarization profile, the same property holds for any other beam obtained from the first one by placing in its path a general linear, deterministic and homogeneous optical element.  In fact, any optical element of this kind can be described by a constant Jones matrix, $\widehat J$, and its effect on a  beam at $z=0$ is to produce at its output the field
\begin{equation}
{\bs V'}_0({\bf r})= 
\widehat J \; {\bs V}_0({\bf r})
\; .
\label{aniso}
\end{equation}
If we introduce the direct propagator in free space, $K_z({\bf r}, {\bs \rho})$, such that
\begin{equation}
{\bs V}_z({\bf r})
= 
\displaystyle\iint
K_z({\bf r}, {\bs \rho})
{\bs V}_0({\bs \rho})
\, {\rm d}{\bs \rho}
\; ,
\label{propDir}
\end{equation}
the expression for the propagated field ${\bs V}'_z$  reads
\begin{equation}
\begin{array}{rl}
{\bs V}'_z({\bf r})
&
=
\displaystyle\iint
K_z({\bf r}, {\bs \rho})
{\bs V}'_0({\bs \rho})
{\rm d}{\bs \rho}
=
\displaystyle\iint
K_z({\bf r}, {\bs \rho})
\, \widehat J \; {\bs V}_0({\bs \rho})
{\rm d}{\bs \rho}
=
\\ \\
&=
\widehat J 
\displaystyle\iint
K_z({\bf r}, {\bs \rho})
{\bs V}_0({\bs \rho})
{\rm d}{\bs \rho}
=
\widehat J \; 
{\bs V}_z({\bf r})
\; ,
\label{spbSecondo}
\end{array}
\end{equation}
so that, if ${\bs V}$ preserves its transverse polarization pattern during propagation, the same occurs for ${\bs V}'$.

Incidentally, we note that the shape-invariance property holds not only when $K_z$ is the direct propagator in free space, but for whatever linear scalar operator. Hence, in particular, it holds for the propagation of the beam through any isotropic optical systems, characterized by an ABCD transformation matrix~\cite{Siegman:Lasers86}.

The best known examples of NUTP fields whose polarization pattern remains unchanged on propagation are donut beams with radial or azimuthal polarization. A way to produce them is through the superposition of a  ${\rm LG}_{0,1}$ mode, with circular polarization ${\hat\epsilon}_1=(\hat x + {\rm i} \, \hat y)/\sqrt{2}$, and a  ${\rm LG}_{0,-1}$ mode, with the same amplitude and orthogonal polarization ${\hat \epsilon}_2=(\hat x - {\rm i} \, \hat y)/\sqrt{2}$ (in both cases $N=1$)~\cite{Gori:JOSAA01,Zhan:AOP09,Ramirez:JOPT09}. An analogous scheme leads to azimuthally polarized beams. Unfortunately, such beams present only linear polarization states across their profile, and this is not enough to perform polarimetric measurements.  In fact, all linear states have $S_3=0$, so that the four independent linear polarization states of the input field cannot be independent.
The use of a homogeneous anisotropic optical element placed before the sample would not help, because even in this case we could not find four independent states across the beam cross section. 

Now we give an example of beams that can be obtained following the above approach. It has been chosen because, as we shall see, it can be well suited for FPP application. Of course, different choices can be made, depending on possible experimental requirements. More and more sophisticated patterns will be obtained on increasing the value of the parameter $N$ and/or the number of involved modes with the same $N$.

For this example we take the following component fields: a single LG mode with indices $p=1$ and $s=3$ (linearly polarized along $\hat x$) and a single LG mode with indices $p=2$ and $s=1$ (linearly polarized along $\hat y$). The sum $2 p + |s|$ gives 5 in both cases. To simplify the notations, in the following we will use normalized coordinates, i.e., we will adopt units for which $w_0$ is unitary.

\begin{figure}[ht]
	\centering
	\includegraphics[width=12cm]{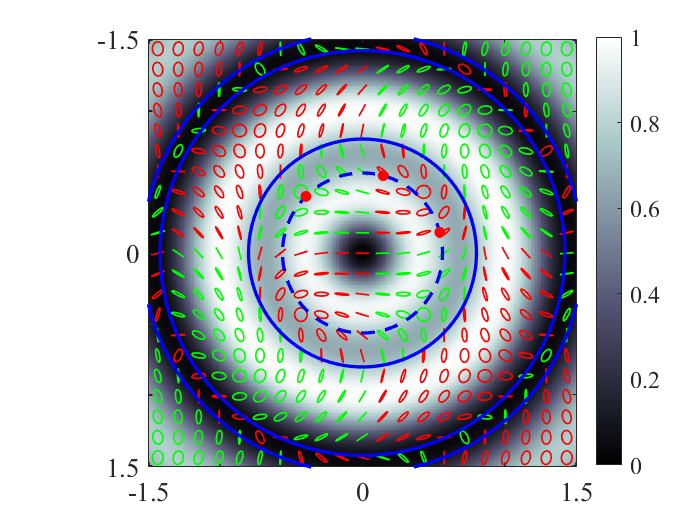}	
	\caption{Transverse intensity and polarization pattern for the field in Eq. (\ref{esempioFP1}).  Handedness is codified by the color of the ellipses (green = right, red = left). The Poincar\'e sphere is mapped twice inside the regions enclosed by solid blue circles (with radii given in Eqs.~(\ref{erre1}) and (\ref{erre})).}
	\label{figFP2}
\end{figure}

Using the definition of LG modes~\cite{Siegman:Lasers86}, the total field can be written as
\begin{equation}
{\bs V}_{0}({\bs r}) 
= 
\left(
\begin{array}{c}
2 r^4 -6 r^2+3
\\ 
\alpha \; r^2 \left( 2 - r^2\right) \; e^{-2 {\rm i} \vartheta}
\end{array}
\right)
r \; e^{- {\rm i} \vartheta} e^{-r^2}
\; ,
\label{esempioFP1}
\end{equation}
where common constant factors have been omitted and the coefficient $\alpha$ (supposed real and positive) has been introduced to let the amplitude ratio between the two fields to be changed. 
The intensity distribution, together with the polarization states of the field across the plane are shown in Fig.~\ref{figFP2}, for $\alpha=1.25 \, \sqrt{2}$. 

It is possible to show that any possible polarization state on the Poincar\'e sphere is present across the transverse plane and even to evaluate the coordinate where it occurs. In fact, from the definition of Stokes parameters ($S_i$, $i=0,...,3$), it turns out that the elements of the normalized Stokes vector ($s_i=S_i/S_0$, $i=1,2,3$) are
\begin{equation}
s_1 = \displaystyle\frac{1-F^2}{1+F^2} \; ;
\hspace{.5cm}
s_2 = \displaystyle\frac{2 F \cos(2 \vartheta)}{1+F^2} \; ;
\hspace{.5cm}
s_3 = \displaystyle\frac{2 F \sin(2 \vartheta)}{1+F^2}
\; ,
\label{esempioFP2}
\end{equation}
where the auxiliary function $F$ is defined as 
\begin{equation}
F = \displaystyle\frac{\alpha \, r^2 \left( 2 - r^2\right)}{2 r^4 -6 r^2+3}
\; .
\label{esempioFP3}
\end{equation}
The latter expression takes all possible values from zero to infinity when $r$ spans the interval $[0, r_1]$, where
\begin{equation}
r_1= \sqrt{\displaystyle\frac{3-\sqrt{3}}{2}}\; 
\label{erre1}
\end{equation}
is the lowest positive root of the denominator of $F$.

The expressions in Eq.~(\ref{esempioFP2}) can be inverted to obtain the polar coordinates where a given polarization (chosen at will) appears across the transverse beam section. For $r\le r_1$ we have:
\begin{equation}
r^2=\dfrac{1+3 \beta-\sqrt{3 \beta^2+3 \beta+1}}{1+2 \beta} \; ,
\label{esempioFP4}
\end{equation}
\begin{equation}
\vartheta = \displaystyle\frac{1}{2} \arg[s_2 + {\rm i} s_3] \;\;\;\;(+ \pi)
\; ,
\label{esempioFP5}
\end{equation}
with
\begin{equation}
\beta=\frac{1}{\alpha}\sqrt{\dfrac{1-s_1}{1+s_1}} \; ,
\label{salpha}
\end{equation}
so that the whole Poincar\'e sphere is mapped twice in the circle $0 \le r \le r_1$, regardless of the value of $\alpha$. It can be shown that the Poincar\'e sphere is mapped again (twice) in the region $r_1 \le r \le r_2$ and in the region  $ r_2\le r \le r_3$, with
\begin{equation}
r_2=\sqrt{2} \; ; \;\;\;  r_3= \sqrt{\displaystyle\frac{3+\sqrt{3}}{2}}\; 
\; .
\label{erre}
\end{equation}
The three regions are identified by the solid blue circles in Fig.~\ref{figFP2}.

Since we are interested in using a FPB as a parallel polarization state generator in FPP, we have to select the four optimum input polarization states and to find the points where they are located. 
It is well known that the optimum configuration under minimum number of testing states in a Mueller matrix polarimeter involves four input states of polarization that form a regular tetrahedron inscribed in the Poincar\'e sphere \cite{Azzam:JOSAA88,Layden:OE12,Twietmeyer:OE08}. 

A useful choice for our aim consists in selecting four points where the intensity values are sufficiently high and, possibly, very close to each other. This is achieved, for instance, on selecting one of the points on the circle of radius $r_1$ (all of them having vertical polarization), then fixing the Stokes vector of one of the tetrahedron vertices (say, $P_0$) at $[-1, 0, 0]^T$.

The three remaining vertices ($P_i$, with $i=1,2,3$) are located on the intersection of the Poincar\'e sphere with the plane $s_1=1/3$. 
For brevity, we introduce the parameter $\sigma=s_2(P_1)$, which can only take values in the interval $-\sqrt{8}/3 \le \sigma \le \sqrt{8}/3$, and assume that the polarization at vertex $P_1$ is left-handed. The value of $\sigma$ fixes the polarization states of all vertices.

It turns out that, limiting ourselves to the region $r \le r_1$, the three polarization states with $s_1=1/3$ are located on a circle having radius 
\begin{equation}
r^2=\dfrac{3+\alpha \sqrt{2}-\sqrt{2\alpha^2+3 \alpha \sqrt{2} +3}}{2+\alpha \sqrt{2}} \; ,
\label{raggioblu2}
\end{equation}
at values of $\theta$ given by
\begin{eqnarray}
\begin{array}{c}
 \cos\left(2 \theta_{1}\right)=\dfrac{3 \sigma}{2\sqrt{2}} \; ; \,\,\, \,\,\,
   \cos\left(2 \theta_{2,3}\right)=\dfrac{3}{2\sqrt{2}} \left(\dfrac{-\sigma}{2}\pm \sqrt{\dfrac{2}{3}-\dfrac{3 \sigma^2}{4}}\right) .  
\end{array}
\end{eqnarray}
Further solutions exist for $r>r_1$, but the intensity is lower there.

The measurement points across the beam section, corresponding to $\sigma=0$ are shown in Fig.~\ref{figFP2} as red dots (together with any point on the smallest red circle).
On varying $\sigma$, the three red points move along the dashed blue circle, whose radius is given by Eq.~(\ref{raggioblu2}), but their intensities are always the same. Furthermore, the intensities at all measurement points are very high (always over 75\% of the maximum).

We point out that, if a different polarization basis is chosen for the component fields (instead of $\hat x, {\hat y}$), a different polarization pattern is obtained, but the four selected points keeps presenting polarization states that are on the vertices of a regular tetrahedron inscribed in the Poincar\'e sphere, and can be profitably used for FPP measurements.

A wider class of fields can be identified if the condition of shape-invariance of the polarization pattern is required only when the propagated field reaches the far zone. 
Using the same approach than in the previous case we require that, in the modal expansion of the component fields [Eqs.~(\ref{sviluppoHGinZ})], only those terms are present for which the phase anomalies take the same value in the far field (i.e., when $z \gg z_R$), up to an integer multiple of $2 \pi$. In this case, while in the near zone the mutual phases among the Gaussian modes generally change, thus modifying also the polarization state of the total field, they align again in the far zone.
From a mathematical point of view,  the phase anomaly of each of the Gaussian modes saturate to the value $(N+1) \pi/2$ in the far field. So, to guarantee shape invariance in the far field it is sufficient that either of the two orthogonally polarized components of the beam could be expressed as the sum of HG (and/or LG) modes with values of $n+m$ (and/or of $2p + |s|$) that only differ for $4 \pi \, j$, with integer $j$ \cite{Cincotti:JPA92}.
The FPB presented by Galvez \emph{et al} some years ago \cite{Galvez:AO12} is the superposition of a pure LG mode with $p=s=0$ ($2p+|s|=0$) and a pure LG mode with $p=1$ and $s=2$ ($2p+|s|=4$) with orthogonal linear polarizations, so that it belongs to this class. Following the approach proposed in \cite{Cincotti:JPA92}, 
an endless number of functions of this class can be easily found without resorting to their expansion in Gaussian modes.

\vspace{.5cm}

In conclusion, a class of NUTP beams whose polarization and intensity transverse profiles remain invariant during propagation, up to a scale factor, has been presented and analyzed. The beams of this class that are also FPB's can be profitably used as parallel polarization state generators for Mueller polarimetry. 
The best way to exploit the potential offered by this type of beams is to use a CCD sensor to detect the polarization states across the beam section. Due to the propagation-invariant feature of the transverse polarization pattern, no imaging optical elements are required before the sensor. On taking (at least) four images of the field profile after suitable configurations of linear polarizers and phase plates \cite{chipman18,Goldstein03} and processing them, the whole maps of the Stokes parameters (without and with the sample, respectively) can be obtained. Therefore, in the evaluation of the Mueller matrix of the sample, all possible input polarization states are available at once. A reduction of the propagated errors can be achieved on averaging the results pertaining to different sets of measurement points, or by increasing the number of points used in each measurement \cite{Layden:OE12,Twietmeyer:OE08}.

\vspace{.5cm}

{\bf \large Funding.} Spanish Ministerio de Econom\'ia y Competitividad project PID2019 104268GB-C21.

\vspace{.5cm}

\bibliographystyle{unsrt}

\end{document}